# Application of light diffraction theory to qualify the downstream light field modulation property of mitigated KDP crystals

Hao Yang[1,2,3], Jian Cheng[1,3], Zhichao Liu[2], Qi Liu[1], Linjie Zhao[1], Chao Tan[1], Jian Wang[2], and Mingjun Chen[1]

[1]State Key Laboratory of Robotics and System, Harbin Institute of Technology, Harbin 150001, China

[2]Research Center of Laser Fusion, China Academy of Engineering Physics, Mianyang 621900, China

[3]These authors contributed equally to this work.

Correspondence:
Dr. Mingjun Chen and Dr. Jian Cheng, State Key Laboratory of Robotics and System, School of Mechatronics Engineering, Harbin Institute of Technology, Harbin 150001, China.
Tel.: +86(0)451-86403252. Fax: +86(0)451-86403252.
*E-mail: chenmj@hit.edu.cn and cheng.826@hit.edu.cn*

## Abstract:

Micro-milling can effectively remove laser damage sites on a KDP (potassium dihydrogen phosphate) surface and then improve the laser damage resistance of the components. However, the repaired KDP surface could cause light propagating turbulence and downstream light intensification with the potential risk to damage downstream optics. In order to analyze the downstream light field modulation caused by Gaussian mitigation pits on KDP crystals, a computational model of the downstream light diffraction based on the angular spectrum theory and the Gaussian repair contour is established. The results show that the phase offset caused by the repaired surface produces a large light field modulation near the rear KDP surface. The modulation generated in the whole downstream light field is greater than that caused by the amplitude change. Therefore, the phase characteristics of the outgoing light could be suggested as a vital research topic for future research on the downstream light field modulation caused by mitigation contours. Significantly, the experimental results on the downstream light intensity distribution have good agreement with the simulation ones, which proves the validity of the established downstream light diffraction model. The phase characterization of the outgoing light is proposed as an evaluation tool in the repair of KDP crystals. The developed analytical method and numerical discrete algorithm could be also applicable in qualifying the repair quality of other optical components applied in high-power laser systems.

# 1. Introduction

Potassium dihydrogen phosphate (KDP) is an important material that is ideal for use as Pockel's cells and frequency multipliers in large laser devices, such as National Ignition Facility, Shenguang Facility and Laser MegaJoule [1-4]. However, under strong laser irradiation, laser damage points on the surface of the KDP crystals will form with dimensions ranging from tens of microns to sub-millimeters [5-7]. These damage points generally manifest as laser ablation, micro-cracks, micro-pits or other forms, which seriously limit the light propagation characteristics and service life of the KDP crystal [8-10]. In order to guarantee the performance of these crystals and prolong their life, the concepts of optics repair and recycling strategy are usually proposed [11,12]. When the laser damage points on the surface of KDP crystal are replaced with a pre-designed smooth contour, the growth of damage on the surface of the crystal is significantly mitigated [13,14].

In recent years, surface repair methods and their effects on laser elements have received considerable interest. At present, common micro-defect repair methods include $CO_2$ laser processing, aqueous wet-etching, short pulse laser ablation and micro-machining. Hrubesh *et al*. [15] verified that the micro-machining method is the most promising one for mitigating the surface laser damage sites on KDP crystals. Then, various shapes of the mitigation pits were designed and fabricated on fused silica surface by Bass *et al*. [16] to improve their laser damage resistance. Yang *et al*. [17] compared the internal light intensifications caused by these different shapes and found that the width-depth ratios greater than 5.3 and 4.3 should be applied to spherical and Gaussian repaired contours, respectively, to achieve optimal effects. In addition to the improved laser damage resistance, Matthews *et al*. [18] found that the mitigated contours may cause the laser damage on the downstream optics element due to the light field modulation. Based on Fresnel diffraction theory, Tournemenne *et al*. [19] developed an analytical model to show how small perturbations, such as digs or scratches, can yield intense light enhancements on the downstream axis. Zheng *et al*.

[20] also discovered that a special configuration of the laser damaged coating causes wave-front modulation, which affects the downstream light field propagation.

However, there are few studies on how surface mitigated contours on KDP crystals affect downstream light field propagation. The special contour of the repaired surface may alter the initial phase and amplitude of the outgoing beam [13,21]. The modifications in amplitude and phase relative to the perfect surface can affect downstream field propagation [22]. The analysis of downstream light field modulation depends on what key factors need addressing. On one hand, it is helpful for us to choose the appropriate calculational results as theoretical criteria for future studies. On the other hand, it provides a direction for the further optimization of the repair process. In addition, compared with traditional diffraction problems, the slope of the Gaussian mitigation contour on the KDP crystal surface varies greatly. Therefore, any model needs to account for the cross-section morphology of the repaired contour, and provide a numerical sampling criterion. At present, in current research, there is still a gap in knowledge on these points.

In order to explore the downstream light field modulation generated by the Gaussian mitigation pits on the crystal surface, we first created a downstream light field diffraction computing model. The latter involves a numerical discrete algorithm that is based on the special morphology of the Gaussian mitigation pit. The next step is an analysis and comparison of the influence of the amplitude attenuation, and phase offset of the outgoing beam, on the downstream light propagation. Finally, our theoretical model is validated by the experiment of detecting the intensity distribution of downstream light field. In this study, the sampling conditions of the Gaussian mitigation pits discrete algorithm are proposed, the light field diffraction model suitable for the Gaussian repair contour is established, and the causes of the repair point's modulation of the downstream light field are explained. In addition, the calculation criteria for the modulation of the downstream light field of KDP crystal surface repair are given. In a word, our research results provide technical support for future repair process optimization.

## 2. Model and Theory

Theoretical research on this topic is mainly based on scalar diffraction theory. Since the morphology of crystal surface repair has a major effect on diffraction calculations, we now offer a description on mitigated Gaussian contours.

By means of micro-machining, regular Gaussian contours with smooth surface can be machined on crystal surface. Figure 1(a) shows the optical micrograph of the Gaussian mitigation pit. The repaired area is similar to the surrounding surface with good smoothness and flatness. Insert image is the local details of the cutter grain. The tool path in the repair area is clear, and the material removal is mainly realized by plastic cutting. The two-dimensional morphology of the cross-section contour at the Gaussian mitigation pit is shown in FIG. 1(b). The overall profile is a standard Gaussian curve. The insert is an enlarged local area of the contour. The machined surface is very smooth for the most part. The surface roughness is less than 50 nm, even in areas where the waist of the Gaussian curve is difficult to process. The local area on the right has irregular protrusions. These are mainly related to the cutting tool milling mode (forward or reverse milling) and the contour shape during the cutter processing [23]. Figure 1(c) shows the surface transmittance of the repaired area. Within the test range from 300 nm to 1200 nm, the transmittance of the repaired surface can reach more than 80%. Different from the slight fluctuation of transmittance value around 355 nm, the transmittance remained almost stable around 532 nm, ~ 91.2%.

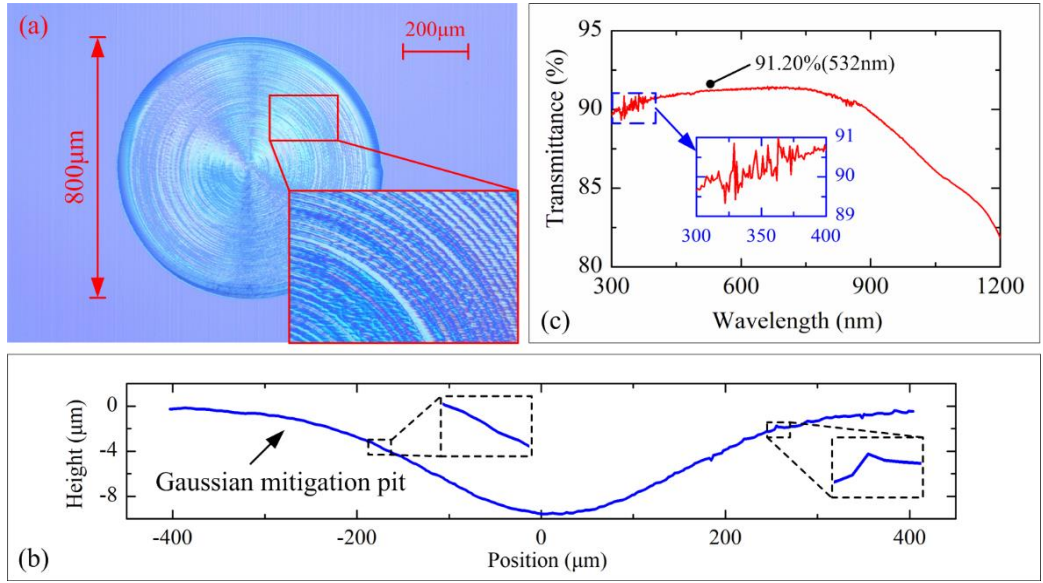

FIG. 1 Gaussian mitigation pit of KDP crystal surface. (a) Optical micrograph of the Gaussian mitigation pit; (b) Two-dimensional morphology of the cross-section contour at the Gaussian mitigation pit; (c) Surface transmittance of mitigation pit.

Figure 2 is a schematic diagram for the theoretical calculation of downstream light field diffraction, for cases where the mitigated KDP crystal is placed in the laser path. The Gaussian mitigation pit is located on the rear surface of the crystal. Light diffraction occurs at a distance $Z$ from the rear surface of the crystal. The electric vector $E$ is regarded as scalar $U(x, y, z)$, which does not include situations where the size of the obstacle (or optics element) is close to the wavelength of the input light [24].

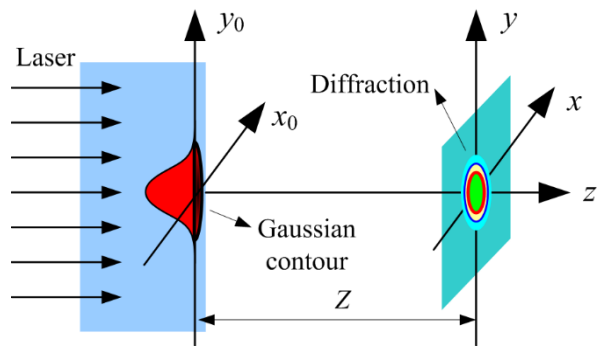

FIG. 2 Schematic diagram for theoretical calculation of downstream light field diffraction behind Gaussian mitigation pit.

When the plane light wave is emitted from the rear KDP surface, a path difference arises in the light due to the effects of the repair depth. Through simple derivation, the phase of the light wave at positions $(x, y, 0)$ on the plane $z = 0$ can be expressed as:

$$\varphi(x, y, 0) = \begin{cases} k(n_{\text{KDP}} - 1) \cdot d_0 \cdot \exp\left(\dfrac{x^2 + y^2}{(w_0/4)^2}\right) & x^2 + y^2 \leq R_0^2 \\ 0 & x^2 + y^2 > R_0^2 \end{cases} \quad (1)$$

where $n_{\text{KDP}}$ is the refractive index of the KDP crystal, $k$ is the wave number of the light, $d_0$ is the maximum depth of the repair area, $w_0$ is the width of the repair area and $R_0 = w_0/2$ is half width of the repair area.

However, due to the reflection of the incident light on the crystal surface, the effects of the microstructure and, especially, the Gaussian contour with different curvatures at different positions, the amplitude of the incident light will decrease. Assuming that the incident light amplitude at the non-repaired area is 1 V/m, the amplitude at positions $(x, y, 0)$ on the plane $z = 0$ can be expressed as:

$$A(x, y, 0) = \begin{cases} \alpha & x^2 + y^2 \leq R_0^2 \\ 1 & x^2 + y^2 > R_0^2 \end{cases} \quad (2)$$

where $\alpha \in (0,1)$ represents the ratio of the amplitude at $x^2 + y^2 \leq R_0^2$ region to other regions.

Then, the complex amplitude at different positions in the plane $z = 0$ can be expressed as:

$$U(x, y, 0) = A(x, y, 0) \cdot \exp[j \cdot \varphi(x, y, 0)] \quad (3)$$

At all passive points, $U$ satisfies the Helmholtz equation, and $U(x, y, 0)$ must be a particular solution to the equation corresponding to $z = 0$. According to the differential equation theory, the Fourier transform and the angular spectrum theory of light propagation, the general solution of the equation can be obtained as follows [25]:

$$U(x, y, Z) = \mathcal{F}^{-1}\left\{\mathcal{F}\{U(x, y, 0)\} \cdot \exp\left[jkZ\sqrt{1 - (\lambda f_x)^2 - (\lambda f_y)^2}\right]\right\} \quad (4)$$

where $F\{\ldots\}$ and $F^{-1}\{\ldots\}$ denote the Fourier transform and the inverse Fourier transform, respectively. The $f_x$ and $f_y$ are the coordinates of the plane wave in the $x$ and $y$ directions of frequency domain, respectively.

We then obtain a relation for the complex amplitude change of the light wave field from $z = 0$ to a point in any downstream plane. The result of light wave propagation in the z-direction is represented in the frequency domain as the spectrum of the initial plane light wave field multiplied by a z-dependent phase delay factor.

For the solution of Eq. (4), the discrete Fourier transform can be solved by computer assistance. With the initial plane sampling width as $L_0$, and the sampling number as $N \times N$, this expression can be rewritten into the following form:

$$U(p\Delta x, q\Delta y) = \text{IFFT}\left\{\text{FFT}\{U_0(m\Delta x_0, n\Delta y_0)\} \cdot \exp\left[jkZ\sqrt{1-(\lambda p\Delta f_x)^2 - (\lambda q\Delta f_y)^2}\right]\right\} \quad (5)$$
$$(p,q,m,n = -N/2, -N/2+1, \cdots, N/2-1)$$

which includes $\Delta x_0 = \Delta y_0 = L_0/N$ as the initial planar discrete Fourier transform, before the corresponding spatial sampling interval. Due to linear system theory, the phase delay factor in Eq. (4) is a transfer function of diffraction in the frequency domain. This shows that the diffraction problem can be viewed as a set of light field waves in a transformation process with a linear space-invariant system. In order to achieve the same coordinate scales, and the product of the operation transfer function, the transfer function in the frequency domain sampling unit must meet the condition $\Delta f_x = \Delta f_y = 1/L_0$. Only at this point, when the product operation is complete and transformed to the space domain, the space domain width of $L = 1/\Delta f_x = 1/\Delta f_y = L_0$ would be restored. According to the angular spectrum theory of light propagation, by reference to the general solution of Eq. (4), for all angular spectrum components that satisfy $1 - (\lambda f_x)^2 - (\lambda f_y)^2 < 0$ will increase exponentially with increasing $Z$, and only those that satisfy $1 - (\lambda f_x)^2 - (\lambda f_y)^2 > 0$, i.e., $f_x^2 + f_y^2 < 1/\lambda^2$ can reach the plane of observation. Therefore, when observing the plane light wave field $U(x, y)$, the highest frequency that may be included in the coordinate direction is:

$$f_{max} = \frac{L_0}{\lambda\sqrt{Z^2 + L_0^2}} \quad (6)$$

In order to obtain satisfactory calculation results, in discrete calculation, the following conditions should be met:

$$\frac{1}{\Delta x} = \frac{1}{\Delta y} = \frac{N}{L_0} \geq 2f_{max} \quad (7)$$

and

$$N \geq \frac{2L_0^2}{\lambda\sqrt{Z^2 + 2L_0^2}} \tag{8}$$

However, since the transfer function in the linear system does not change the spectral width of the initial planar band-limited function $U(x, y, 0)$. When Eq. (3) satisfies Shannon's sampling theorem, the calculated result must also be a diffraction field that satisfies Shannon's sampling theorem.

The amplitudes and phase of the initial plane change with the coordinate plane position $(x, y)$. We focus on the changes to study the mitigation pit light transmittance and structure parameters on downstream light field modulation, respectively. When the planar light wave passes through the repair area, the light is completely obscured in the theoretical model, namely the amplitude of the area of $\alpha = 0$. The downstream light field diffraction due to amplitude variation is defined as amplitude-type diffraction (AD). Similarly, it is assumed that when the light wave passes through the repaired region, the light passes through without any change in amplitude. However, due to the light path difference, the phase offset of the outgoing beam from the repaired contour region does occur. The downstream light field diffraction due to this phase offset is defined as phase-type diffraction (PD).

For the initial plane complex amplitude function of AD, its spatial change rate is small, and the sampling requirement can be met as long as the sample number satisfies Eq. (8). However, for the initial plane complex amplitude function of PD, the spatial rate of change reaches its maximum near $x = y = R_0/4$. The complex amplitude actual phase of the light wave at $z = 0$ and the wrapping phase used in the simulation calculation are shown in Fig. 3(a) and 3(b), respectively. The actual phase of the emitting laser changes continuously in the Gaussian shape. The wrapped phase, calculated from the real and imaginary parts of the field, is limited to $[-\pi, \pi]$. Figure 3(c) shows the comparison between the actual phase and the wrapped phase cross-section curves.

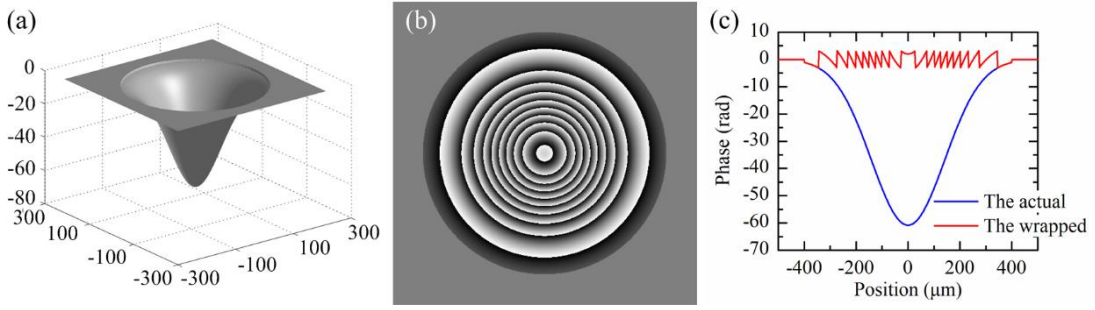

FIG. 3 Initial phase of the outgoing beam on the Gaussian mitigated surface in the theoretical model. (a) The actual initial phase of the complex amplitude of the outgoing beam in the contour area, repaired by the Gaussian pits on the crystal surface; (b) Discontinuous wrapping phase calculated by complex amplitude; (c) Comparison of actual phase and wrapped phase section curves.

In order to avoid initial light-wave distortion, the reciprocal of sampling spacing $1/\Delta x_0$ need to be not less than twice the highest frequency of function. There are at least two sampling points in the period of space corresponding to the highest frequency where:

$$2 \cdot \left| \frac{\partial}{\partial x} \cdot \frac{2\pi(n-1)}{\lambda} \cdot d_0 \cdot \exp\frac{-(x^2+y^2)}{(R_0/2)^2} \right|_{x,y=\frac{\sqrt{2}}{4}R_0} \cdot \frac{L_0}{N} \leq 2\pi \qquad (9)$$

That is:

$$N \geq \frac{4\sqrt{2}d_0 L_0(n-1)}{R_0 \lambda e} \qquad (10)$$

The above are the sampling conditions that need to be satisfied for the diffraction model for Gaussian mitigation pits on KDP surface.

The modulation $M$ is defined as the ratio between the maximum light intensity on the diffraction plane and the initial incident light intensity at different positions on the downstream light field from the rear KDP surface. As the light intensity at the spatial position $I \propto |E|^2 = U \cdot U^*$, the modulation can be obtained by the complex amplitude function.

## 3. Experiment

In this section, the experimental method for directly detecting the downstream light intensity, behind the mitigated crystal surface, is used to verify the theoretical calculations. First, the Gaussian mitigation pits should be machined on the crystal surface. The specific processing methods have been described in previous articles [17].

Then, the mitigation pit needs to be exposed to a uniform, collimated laser in order to capture and analyze the diffraction image downstream in the detecting plane. Figure 4 shows the detection platform for the light intensity distribution of the downstream diffraction light field. We use a continuous laser with an output wavelength of 532 nm. The laser passes through the mirror to change the direction of propagation, and the laser output is controlled by the shutter. The laser is then spatially filtered using an aperture and collimated using a glass lens. After collimating, the laser is irradiated onto the mitigated KDP surface. The effective laser spot diameter is ~5 mm, which is much larger than the width of the repair area. Finally, the spatial light intensity distribution is detected by CCD for subsequent data analysis.

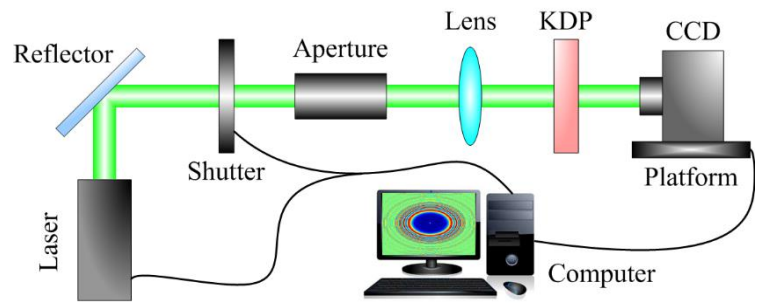

FIG. 4 Detection platform for light intensity distribution of downstream diffraction light field.

## 4. Results and discussion

**4.1 Downstream light field modulation generated by amplitude-type diffraction**

Using the discrete algorithm of angular spectrum diffraction theory and the numerical sampling method based on Gaussian repair geometric contour, the intensity distribution of different downstream locations obtained by AD is calculated. When the diffraction distance is 5 mm, the change of beam amplitude causes the diffraction pattern center at this position to have almost no light intensity. The light field distribution around the repaired contour is more uniform. When the diffraction distance is 100 mm, an on-axis hot-spot appears at the center of the diffraction pattern. There are diffraction rings around the hot-spots on the pattern. When the diffraction distance is 500 mm, the central hot-spot occupies a larger space than the diffraction distance of 100 mm. The number of bright rings increased. When the diffraction distance continues to increase to 1000 mm, the surface excitation of the central hot-spot and the radius of the surrounding bright ring continue to increase, but the light intensity gradually decreases. The calculated results are similar to those of disk diffraction [26]. At this point, the repaired contour plays a role of circular obscuration in the light field diffraction. The AD can be considered as a special case of disk diffraction.

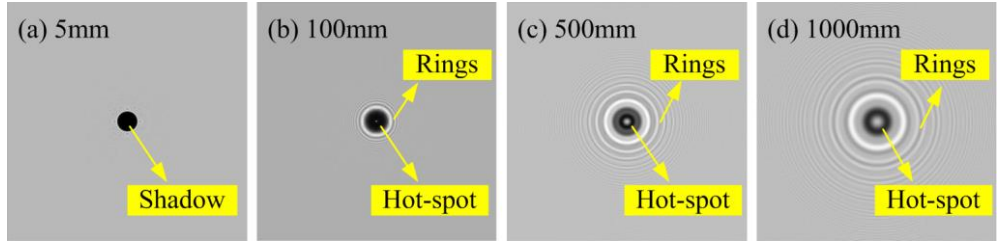

FIG. 5 Light intensity distribution at different downstream locations from the rear KDP surface obtained by AD: (a) 5 mm; (b) 100 mm; (c) 500 mm; (d) 1000 mm.

**4.2 Downstream light field modulation generated by phase-type diffraction**

When the PD numerical calculation meets the numerical sampling method for Gaussian repair geometric contour, the light intensity distribution at different downstream locations is shown in Fig. 6. When the diffraction distance is small, the PD calculation results are very different from those of AD. The repair contour edge generates high intensity off-axis bright ring at $Z = 5$ mm. As $Z$ increases to 100 mm, the off-axis bright ring radius gradually increases, and its brightness is greater than that of AD calculated at this position. When the diffraction distance $Z$ is greater than 500 mm, the PD pattern is similar to the AD pattern. However, the size and brightness of the bright spot in the middle of the PD pattern are smaller than that of the AD pattern. AD and PD are two extreme cases of actual diffraction in the downstream light field behind the repaired surface. Therefore, increasing the proportion of PD in the actual diffraction can reduce the size of the on-axis hot-spot at the far location of the diffraction.

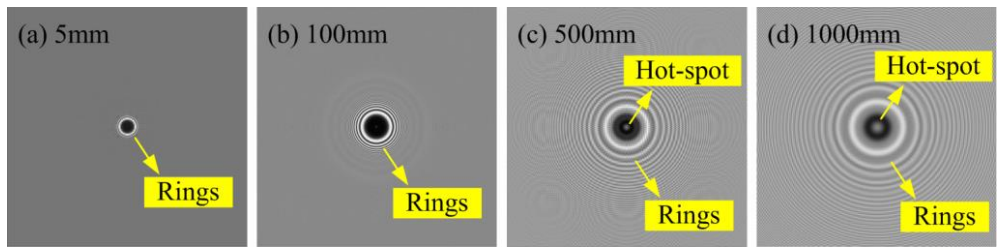

FIG. 6 Light intensity distribution at different downstream locations from the rear KDP surface obtained by PD: (a) 5 mm; (b) 100 mm; (c) 500 mm; (d) 1000 mm.

## 4.3 Comparison between amplitude-type diffraction and phase-type diffraction

In order to analyze the differences between AD and PD on downstream light field modulation, the calculated data for the downstream light field modulation are compared. In Fig. 7(a), the maximum modulation curve of PD (the red) is always at the highest position among the four curves. The curves rise rapidly after a small drop at the initial position. When the modulation curve rises to ~ 9, the curve begins a steep descent, producing a modulation peak. Then the curve flattens out as the diffraction distance increases, and the modulation stabilizes near 2. The maximum light field modulation curve generated by AD (the blue) is located below that of PD. At the position close to the rear KDP surface, the change in trend of these two is opposite with the diffraction distance. As shown in Fig. 7(a) insert, the peak value of the blue curve corresponds to the minimum of the red curve. However, as the diffraction distance increases, the blue curve does not fluctuate significantly like the red curve, and the corresponding modulation is almost stable around 1.5. The AD on-axis modulation curve (the orange) and the PD on-axis modulation curve (the green) are located at the bottom and interwoven. The overall trend of the green curve is the opposite of that of the red, but the corresponding modulation range is much smaller. The orange and the green curve have a similar change trend, but when the diffraction distance is less than 35 mm, the corresponding modulation of the orange curve is ~ 0. When the diffraction distance is less than 80 mm or greater than 220 mm, the green curve modulation has significant noise. The orange curve is always smooth.

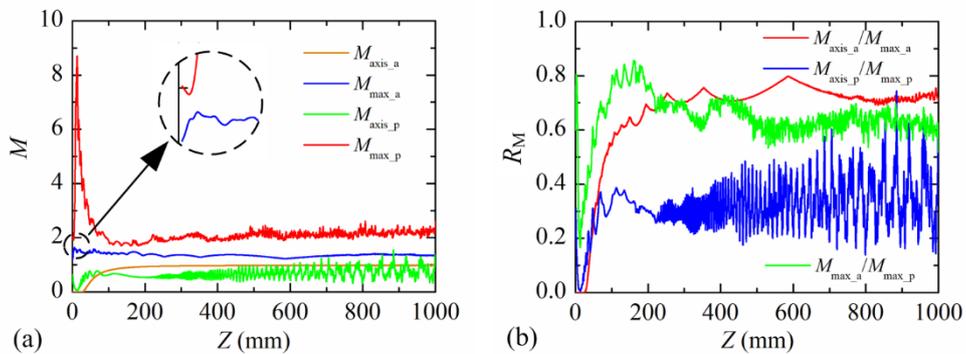

FIG. 7 Comparison of calculated results of AD and PD. (a) Curves of downstream light field modulation on axis and maximum modulation due to AD and PD (b) Modulation ratio curves: AD and

PD on-axis modulation occupies each maximum modulation ratio curve, and AD maximum modulation occupies the PD maximum modulation ratio curve.

Data post-processing is performed on the downstream light field modulation obtained by AD and PD calculations, and the modulation ratio curve is drawn as shown in Fig. 7(b). The red curve (representing $M_{axis\_a}/M_{max\_a}$) coincides with the abscissa axis at the initial position as the AD on-axis modulation near the crystal surface is almost 0. Removing this part of the curve, the red curve has the same trend as the blue curve (representing $M_{aixs\_p}/M_{max\_p}$). When Z > 50 mm, the curves increase rapidly to the higher position. Corresponding to Fig. 7(a), the on-axis modulation and the maximum modulation curves obtained by calculation of the same diffraction type tend to be close to each other as the diffraction distance increases. This indicates that with the increase of the diffraction distance, the concentration effect at the center of the diffraction field plays a more and more important role in the light field modulation. This is similar to the previous research results for fused silica [18,27]. The mitigated contour effect of a convex lens causes an increase in the light intensity at the center of the far field. However, the om-axis modulation is always less than the maximum modulation. This is different from the effect of fused silica mitigation pits on the downstream light field modulation. The maximum light intensity enhancement position is always located in the surrounding bright ring, which is the main factor controlling the light field modulation. The blue curve is almost always below the red curve, indicating that the central hot-spot has a much greater influence on light propagation in AD than in PD. The green curve represents the ratio of AD maximum modulation to PD maximum modulation. The curve falls steeply, rises rapidly, then falls slightly and remains high. When the diffraction distance is 1000 mm, the maximum modulation of PD is still larger than that of AD. PD does not affect the energy of the whole light field system, but only changes the spatial phase distribution of the beam; this is why a large number of light intensifications is shown in PD. It seems that increasing the proportion of AD in the diffraction could reduce downstream field modulation. However, the transmittance of the repaired surface is generally related to the quality of the finished surface. Reducing the surface quality of the mitigation pits adversely affects the laser damage resistance of the component itself. Although reducing the adverse effect of the mitigation pits on the optics itself and on the downstream element seems contrary, we can minimize the laser damage threat to the downstream element by finding the appropriate installation location. In addition, the PD calculations of the downstream light field modulation are always the upper limit of actual diffraction light field modulation. Therefore, the calculation results of PD are selected as a strict standard for the later analysis of downstream light field modulation.

**4.4 Experimental results of detecting downstream light intensity distribution**

By using the detection platform for the light intensity distribution of the downstream diffraction light field, we obtain the light intensity distribution at the position of 500 mm from the rear KDP surface (Fig. 8(a)). The detected diffraction pattern is generally distributed in concentric circles. A light intensity hot-spot appears at the center. This is similar to the calculated diffraction pattern at $Z = 500$ mm in Section 4.1 and 4.2. However, there are some irregular bright spots in the upper left corner of the bright ring around the experimental diffraction pattern. Based on the locations of these bright spots and the repair morphology in Fig. 1, we infer that these bright spots are caused by the irregular protrusions on the repair surface. Micro-machining on KDP crystal uses a high-speed ball-end milling cutter to remove the damage material. The milling cutter presents different milling modes in different areas along the horizontal direction of the repaired contour, such as forward milling, reverse milling or mixed forward-reverse milling. In addition, the effective cutting speed of the cutter is variable in different positions along the vertical direction of the repaired contour, because the machining point on the ball-end milling cutter changes with the slope of the repaired contour. Thus, the quality of the repaired surface depends on the region of the repaired surface. In Fig. 1(c) on the right side of the cross-section profile, near the waist of the Gaussian curve, the surface mass is relatively low and it is accompanied by protrusions. We speculate that this is the cause of the abnormal bright spots in the diffraction pattern. Therefore, the reduction of the influence of the milling mode and its cutting speed is particularly important in future research on the repair technology of spherical cutters with complex curved surface. Fig. 8(b) shows the calculation of AD and PD, and the experimental results. The three curves show a similar trend, in the shape of the letter W. The PD results are the maximum light intensity. This is consistent with the simulation results. Although the light intensity distribution of the experiment is similar to the trends of the simulation results, the light intensity at the edge position is slightly lower than the calculated results of AD. It is speculated that this is related to the attenuation of the light caused by the air. In general, the simulation results are verified by the experiment.

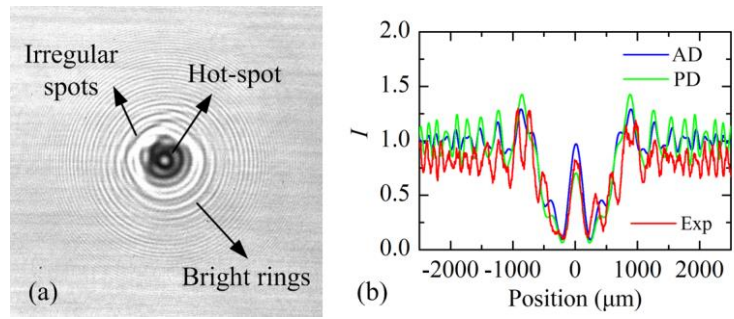

FIG. 8 Comparison of the simulations and the experiments. (a) Detection results of downstream light intensity distribution. (b) Calculations of AD, PD and the experimental results.

## 5. Conclusion

Experimental and theoretical methods were used to explore the causes of downstream light field modulation generated by Gaussian mitigation pits on KDP surfaces. Based on the Gaussian-profile characteristics and scalar diffraction theory, a computational model of the downstream light field modulation was established. The downstream light field modulation generated by AD and PD was simulated. The results show that the downstream light field modulation is caused by the amplitude and phase change of the outgoing beam in the repaired region. However, the modulation caused by PD is more intense, especially near the crystal repair surface. When the diffraction distance exceeds 100 mm, the modulation effects caused by AD and PD are similar. This indicates that the diffraction depends more on the phase offset of the outgoing beam in positions near the mitigation pits. When the diffraction distance is far enough, the phase offset and amplitude attenuation contribute little to the light propagation, but the PD is slightly dominant. The experimental results on the downstream light intensity distribution agree well with the simulations, which proves the validity of the established downstream light diffraction model. In addition, the calculation results of PD light field modulation are the upper limit of the actual modulation of downstream light field diffraction. In the future work of this field, the method of PD calculation can be used as a research tool for optimizing the processing technology and structural parameters of mitigation pits. The calculation model of downstream light field diffraction based on the special contour will be also very useful for analyzing and qualifying the downstream modulation property of surface microstructures on other transparent materials.

## Acknowledgements

This work was supported by Science Challenge Project (No. TZ2016006-0503-01); National Natural Science Foundation of China (Nos. 51775147, 51705105); Young Elite Scientists Sponsorship Program by CAST (No. 2018QNRC001); China Postdoctoral Science Foundation (Nos. 2017M621260, 2018T110288); Heilongjiang Postdoctoral Fund (No. LBH-Z17090); Self-Planned Task Foundation of State Key Laboratory of Robotics and System (HIT) of China (Nos. SKLRS201718A, SKLRS201803B); Southwest University of Science and Technology (19kfzk03).